\documentclass[pre,10pt,twocolumn,showpacs,preprintnumbers,amsmath,aps]{revtex4-1}
\usepackage{graphicx} 
\usepackage{bm}
\usepackage{subfigure}

\def\nbC{{\mathchoice {\setbox0=\hbox{$\displaystyle\rm C$}%
\hbox{\hbox to0pt{\kern0.4\wd0\vrule height0.9\ht0\hss}\box0}}
{\setbox0=\hbox{$\textstyle\rm C$}\hbox{\hbox
to0pt{\kern0.4\wd0\vrule height0.9\ht0\hss}\box0}}
{\setbox0=\hbox{$\scriptstyle\rm C$}\hbox{\hbox
to0pt{\kern0.4\wd0\vrule height0.9\ht0\hss}\box0}}
{\setbox0=\hbox{$\scriptscriptstyle\rm C$}\hbox{\hbox
to0pt{\kern0.4\wd0\vrule height0.9\ht0\hss}\box0}}}}
\def\nbQ{{\mathchoice {\setbox0=\hbox{$\displaystyle\rm
Q$}\hbox{\raise
0.15\ht0\hbox to0pt{\kern0.4\wd0\vrule height0.8\ht0\hss}\box0}}
{\setbox0=\hbox{$\textstyle\rm Q$}\hbox{\raise
0.15\ht0\hbox to0pt{\kern0.4\wd0\vrule height0.8\ht0\hss}\box0}}
{\setbox0=\hbox{$\scriptstyle\rm Q$}\hbox{\raise
0.15\ht0\hbox to0pt{\kern0.4\wd0\vrule height0.7\ht0\hss}\box0}}
{\setbox0=\hbox{$\scriptscriptstyle\rm Q$}\hbox{\raise
0.15\ht0\hbox to0pt{\kern0.4\wd0\vrule height0.7\ht0\hss}\box0}}}}
\def\nbT{{\mathchoice {\setbox0=\hbox{$\displaystyle\rm
T$}\hbox{\hbox to0pt{\kern0.3\wd0\vrule height0.9\ht0\hss}\box0}}
{\setbox0=\hbox{$\textstyle\rm T$}\hbox{\hbox
to0pt{\kern0.3\wd0\vrule height0.9\ht0\hss}\box0}}
{\setbox0=\hbox{$\scriptstyle\rm T$}\hbox{\hbox
to0pt{\kern0.3\wd0\vrule height0.9\ht0\hss}\box0}}
{\setbox0=\hbox{$\scriptscriptstyle\rm T$}\hbox{\hbox
to0pt{\kern0.3\wd0\vrule height0.9\ht0\hss}\box0}}}}
\def\nbS{{\mathchoice
{\setbox0=\hbox{$\displaystyle     \rm S$}\hbox{\raise0.5\ht0%
\hbox to0pt{\kern0.35\wd0\vrule height0.45\ht0\hss}\hbox
to0pt{\kern0.55\wd0\vrule height0.5\ht0\hss}\box0}}
{\setbox0=\hbox{$\textstyle        \rm S$}\hbox{\raise0.5\ht0%
\hbox to0pt{\kern0.35\wd0\vrule height0.45\ht0\hss}\hbox
to0pt{\kern0.55\wd0\vrule height0.5\ht0\hss}\box0}}
{\setbox0=\hbox{$\scriptstyle      \rm S$}\hbox{\raise0.5\ht0%
\hboxto0pt{\kern0.35\wd0\vrule height0.45\ht0\hss}\raise0.05\ht0%
\hbox to0pt{\kern0.5\wd0\vrule height0.45\ht0\hss}\box0}}
{\setbox0=\hbox{$\scriptscriptstyle\rm S$}\hbox{\raise0.5\ht0%
\hboxto0pt{\kern0.4\wd0\vrule height0.45\ht0\hss}\raise0.05\ht0%
\hbox to0pt{\kern0.55\wd0\vrule height0.45\ht0\hss}\box0}}}}
\def\nbZ{{\mathchoice {\hbox{$\sf\textstyle Z\kern-0.4em Z$}}
{\hbox{$\sf\textstyle Z\kern-0.4em Z$}}
{\hbox{$\sf\scriptstyle Z\kern-0.3em Z$}}
{\hbox{$\sf\scriptscriptstyle Z\kern-0.2em Z$}}}}

\begin{document}

\title{Decorrelation of the static and dynamic length scales in hard-sphere glass-formers}

\author{Patrick~Charbonneau}\email{patrick.charbonneau@duke.edu}
\affiliation{Department of Chemistry, Duke University, Durham,
North Carolina 27708, USA}
\affiliation{Department of Physics, Duke University, Durham,
North Carolina 27708, USA}
\author{Gilles Tarjus} \email{tarjus@lptl.jussieu.fr}
\affiliation{LPTMC, CNRS-UMR 7600, Universit\'e Pierre et Marie Curie,
bo\^ite 121, 4 Pl. Jussieu, 75252 Paris c\'edex 05, France}

\date{\today}

\begin{abstract}
We show that in the equilibrium phase of glass-forming hard-sphere fluids in 
three dimensions, the static length scales tentatively associated with the dynamical slowdown and the 
dynamical length characterizing spatial heterogeneities in the dynamics unambiguously decorrelate. 
The former grow at a much slower rate than the latter when density increases. This observation is valid for the 
dynamical range that is accessible to computer simulations, which roughly corresponds to that accessible in colloidal experiments. We also find 
that in this same range, no one-to-one correspondence between relaxation time and point-to-set correlation length exists.  
These results point to the coexistence of several relaxation mechanisms in the dynamically accessible 
regime of three-dimensional hard-sphere glass formers.
\end{abstract}


\maketitle

\section{Introduction}

A recurring question about  glass formation concerns the collective nature of the dynamics 
as one cools or compresses a liquid. If the phenomenon is collective, it should be 
characterized by the development of nontrivial correlations to which one 
or several typical length scales might be associated. One source of growing correlations has been clearly 
identified in connection with the increasingly heterogeneous character of the dynamics as the 
system becomes more sluggish. An associated length, commonly referred to as ``dynamical'', can 
then be extracted from multi-point space-time correlation functions~\cite{kirkpatrick:1988,franz:1999,berthier:2005,
bouchaud:2005,berthier:2011b,footnote:0}. In addition, several theories of 
the glass transition posit the existence of a growing ``static'' 
length accompanying a liquid's dynamical slowdown. This length is, however, undetectable 
through standard measurements of pair density correlations, which have been shown to 
display but unremarkable changes as the structural relaxation slows down.

Proposals for unveiling such a nontrivial static length include measures of the spatial extension of 
some locally preferred structure, as obtained from static correlations of a  
bond-orientational order parameter~\cite{steinhardt:1983,ernst:1991,kawasaki:2007,shintani:2008,watanabe:2008,
leocmach:2012,sausset:2008d,sausset:2010,malins:2012,xu:2012,xu:2012b} or, via dimensional analysis, from the occurrence 
frequency of a given local arrangement~\cite{dzugutov:2002,miracle:2004,anikeenko:2007,coslovich:2007,coslovich:2011,
aharonov:2007,pedersen:2010,karmakar:2012,speck:2012,soklaski:2013}. Such proposals have long been advocated, 
but their usefulness remains uncertain~\cite{charbonneau:2012,charbonneau:2012b}. More recently, approaches that detect the 
growth in static correlations while staying clear of 
any specific proposal about local order, \textit{i.e.},  ``order-agnostic'' approaches, have been developed. 
Among these proposals, we note patch repetition lengths~\cite{kurchan:2011,sausset:2011}, 
length scales extracted from information theoretic analysis~\cite{ronhovde:2011,dunleavy:2012} 
or from finite-size studies of the configurational entropy~\cite{karmakar:2009}, and other ``point-to-set'' correlation 
lengths~\cite{bouchaud:2004,montanari:2006,biroli:2008,berthier:2012,hocky:2012,kob:2012}.

Point-to-set correlations play a special role in the theory of the glass transition. They are more general than structural lengths based on a specific 
local-order description, and are expected to provide upper bounds for the latter. As further 
discussed below, a type of point-to-set correlation length even enters in an upper bound for the relaxation time of the liquid~\cite{montanari:2006}.
Point-to-set correlation lengths can be studied by considering the distance over which boundary conditions imposed 
by pinning particles in a liquid configuration affect the equilibrium structure of the remaining (unpinned) particles. 
The original proposal, motivated by the random first-order transition theory~\cite{kirkpatrick:1989}, considered 
a cavity whose exterior is a frozen liquid configuration~\cite{bouchaud:2004}. 
Other geometries of the set of pinned particles 
also allow one to extract point-to-set correlation lengths~\cite{berthier:2012,cammarota:2011,cammarota:2013}, although the lengths measured for different geometries
need not coincide nor evolve in exactly the same way as temperature decreases or density increases~\cite{jack:2012}. 

It has also been suggested that a length scale could be obtained from the finite-size analysis of the relaxation time 
itself, in a finite system with either periodic boundary conditions~\cite{karmakar:2009,berthier:2012,berthier:2012b} or with a pinned wall 
boundary~\cite{scheidler:2004,berthier:2012,kob:2012}. Such a length has been 
called ``dynamical'' in Refs.~\onlinecite{berthier:2012,kob:2012}, but it should be kept in mind that it is \textit{a priori} 
different from the length characterizing the extent of the dynamical heterogeneity~\cite{flenner:2012}.

It is worth noting that, at present, none of the aforementioned lengths are directly accessible in experimental
glass-forming liquids. The situation is slightly better for colloids and granular materials, but most of the information 
on these lengths must still be obtained from model, yet realistic, glass formers via computer simulations.

Given this panorama of length scales that appear in the context of glass formation, many questions 
can be raised, among which the following two are central to the glass problem.

(i) \textit{Is the temperature (or pressure) evolution of these various lengths correlated?}
 
In other words, to which extent are the static lengths correlated amongst themselves, the 
dynamical lengths correlated amongst themselves, and the static and dynamical lengths correlated with each other? Due to the limited growth 
of the static lengths that is generally observed, it seems hard to decide on the first issue. For the second, most data on 
dynamical lengths characterize the typical extent of the heterogeneous dynamics and are consistent with each other. 
As mentioned above, there is evidence 
that a ``dynamical'' length extracted from finite-size scaling behaves differently~\cite{kob:2012,flenner:2012}, but we shall 
here mainly focus on the lengths extracted from four-point space-time correlation functions. 

The third issue is more contentious. It has been forcefully advocated by Tanaka and 
coworkers~\cite{kawasaki:2007,shintani:2008,watanabe:2008,leocmach:2012} that 
the structural length extracted from 
the (static) correlations in a bond-orientational order parameter and 
the dynamical length obtained from four-point space-time correlation functions perfectly correlate. A somewhat different 
result has, however, been obtained by 
Sausset et al. in their study of a glass-forming liquid 
on the hyperbolic plane~\cite{sausset:2010}, where the convergence of the two types of lengths 
is found to depend on the dynamical regime under consideration. 
A conclusion also at odds with that of Tanaka and coworkers has recently been reached by 
Xu \emph{et al.}~\cite{xu:2012b} for a polydisperse two-dimensional Lennard-Jones mixture  
and by Dunleavy \emph{et al.}~\cite{dunleavy:2012} for binary three-dimensional hard-sphere mixtures.

(ii) \textit{Is the increase of the relaxation time due to the growth of any of the above lengths, or, with less compelling 
consequences, is it at least correlated to it?}
 
Guided by known forms of dynamical scaling near critical points, several relations between relaxation time 
and lengths have been suggested and tested, such as a conventional power-law, $\tau_\alpha \sim \xi^z$, 
and activated scaling behavior, $\log(\tau_\alpha) \sim \xi^{\psi}$, with prefactors that possibly 
depend on temperature and pressure. Empirical correlations 
of one sort or the other have indeed been found with either the 
conventional scaling and a dynamic length~\cite{lacevic:2003,whitelam:2004,berthier:2007b,flenner:2009,kim:2012}  
(with $z$ varying from 2 to 5) or the activated expression and a dynamical~\cite{capaccioli:2008,flenner:2011,flenner:2012} 
or a static length~\cite{kawasaki:2007,shintani:2008,watanabe:2008,karmakar:2012,sausset:2010,hocky:2012} 
(with $\psi$ roughly varying from $1$ to $2$). 

Yet, as stressed by Harrowell and coworkers~\cite{widmer-cooper:2004,widmer-cooper:2006,widmer-cooper:2009}, 
correlation does not imply causation. The fundamental question to be addressed is therefore whether 
one can find a causal link between the increase of the relaxation time and that of any of the 
proposed length scales. An important result in this direction has been obtained 
by Montanari and Semerjian~\cite{montanari:2006}: the relaxation time is bounded from above by an activated-like 
formula involving a static point-to-set correlation length $\xi_{\rm PS}$,
\begin{equation}
\label{eq_bound_tau}
\tau_\alpha \leq \tau_{0}\exp\left (B\, \xi_{\rm PS}^d \right),
\end{equation}
where $\tau_{0}$ sets the microscopic time scale and $d$ is the spatial dimension. The coefficient $B$ depends on temperature and 
pressure, and is such that when $\xi_{\rm PS}$ is about one particle size the right-hand side describes the 
``noncooperative dynamics'' of the model~\cite{franz:2011}. According to the above equation, the relaxation time $\tau_{\alpha}$ thus cannot 
diverge at a finite temperature nor at a finite pressure without the concomitant divergence of a static length.

In this article, we address the above two questions by considering three-dimensional glass-forming hard-sphere mixtures. A point-to-set length  
has been obtained through the random pinning of a set of particles in an equilibrated configuration. Results have already been shown in 
Ref.~[\onlinecite{charbonneau:2012c}] and are complemented here by additional computations and a slightly improved methodology. We have also calculated 
a dynamical length via a four-point space-time correlation function. In conjunction with our previous investigation of structural lengths 
associated with local order~\cite{charbonneau:2012,charbonneau:2012c},  this study then allows us to unambiguously conclude 
that the evolution with pressure (or packing fraction) of the static lengths decorrelate from that of the dynamical length in 
the range of relaxation times accessible to computer simulations. Additionally, no one-to-one correspondence  between 
relaxation time and static length(s) is found. 

\section{Models and method}

\subsection{Model and simulation}
\begin{figure}
\includegraphics[width=\columnwidth]{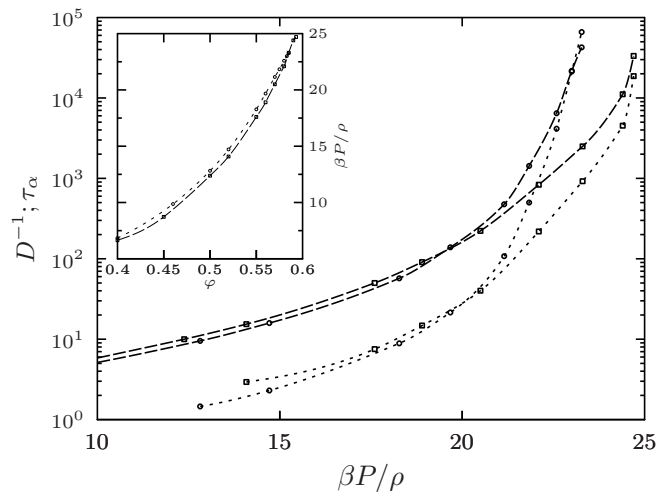}
\caption{Relaxation time (dotted line) and diffusivity (dashed line) of large spheres in two equimolar binary 
hard-sphere glass formers (7:5 squares; 6:5 circles) versus the reduced pressure $\beta P/\rho$, where $\rho=N/V$ is the number density. 
Lines are guide for the eye. (Inset) Equation of state for the two mixtures.} 
\label{fig:relaxtime}
\end{figure}

\subsection{Static lengths} 
\begin{figure}
\includegraphics[width=\columnwidth]{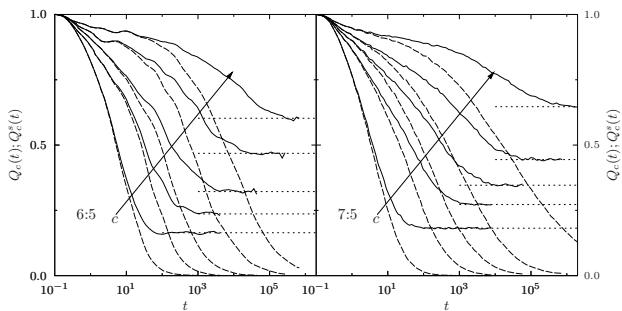}
\caption{Time evolution of the overlap $Q_c$ (solid line) and its self-component $Q_c^s$ (long-dashed line) for the two glass-forming hard-sphere mixtures at a 
packing fraction $\varphi=0.55$ for $c\approx$ 1\%, 6\%, 10\%, 15\%, and 20\% (6:5 mixture) and 1\%, 10\%, 15\%, 20\%, 30\% (7:5 mixture). The asymptotic value of $Q_c$ (short-dashed line) is reached when  $Q_c^s$ has completely decayed.}
\label{fig:overlap}
\end{figure}

\begin{figure}
\includegraphics[width=\columnwidth]{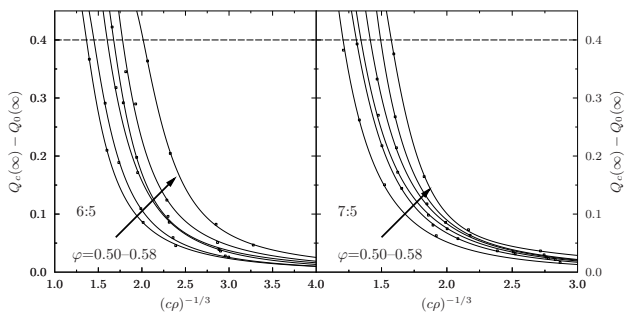}
\caption{Variation of the asymptotic value of the overlap difference $Q_c(\infty)-Q_0(\infty)$ 
with increasing distance between pinned particles $(c \rho)^{-1/3}$ for $\varphi=$ 0.50, 0.52 0.55, 0.56, 0.57, and 0.58. Solid lines are polynomial fits to the numerical results. As the packing fraction increases, the crossover from 
high to low overlap takes place at a longer length.}
\label{fig:pinning}
\end{figure}

We consider two glass-forming hard-sphere systems in $d$=3. The equimolar 
binary mixtures of spheres with diameter ratios $\sigma_1$:$\sigma_2$ of 7:5 and 6:5 ($\sigma_1$ sets 
the unit length) are selected to prevent crystallization. The properties of these mixtures 
have been extensively characterized~\cite{charbonneau:2012c}, notably in Refs.~\cite{berthier:2009,flenner:2011} 
and Refs.~\cite{foffi:2003,foffi:2004}, respectively. Equilibrated fluid configurations 
over a range of packing fractions $\varphi$ with at least $N=1236$ particles for the static length and $N=79,104$ for the 
dynamical length are obtained under periodic boundary conditions using 
a modified version of the event-driven molecular dynamics code described in 
Refs.~\cite{skoge:2006,charbonneau:2012b}.  Quantities are obtained from averaging between 4 and 8 independent replicates for each system. Time is expressed in units of 
$\sqrt{\beta m\sigma_1^2}$ for particles of unit mass $m$ at fixed unit inverse temperature 
$\beta$. The diffusivity $D$ is obtained by measuring the long-time behavior of the mean-squared 
displacement $\lim_{t\rightarrow\infty}\langle (\Delta r)^2\rangle(\equiv\frac{1}{N}\sum_i [\mathbf{r}_i(t)-\mathbf{r}_i(0)]^2)=2dDt$, 
the pressure $P$ is mechanically extracted from the collision statistics, and 
the structural relaxation time $\tau_{\alpha}$ is measured as explained in Sect.~\ref{sect:dynamical} (Fig.~\ref{fig:relaxtime}). We now detail the 
procedures for extracting the static and dynamical lengths.

To compute a static point-to-set correlation length, one may consider a system in which 
a fraction $c$ of the particles of an equilibrium hard-sphere fluid configuration are pinned at random. 
Information on point-to-set correlations in the (bulk) fluid is then 
obtained from the long-time limit of the overlap  between the original configuration and the configuration equilibrated in 
the presence of the pinned particles. If the reference and the final configurations are quite similar, the average pinning spacing 
is shorter than the static correlation length, and the opposite is true if the two configurations are dissimilar. To measure 
the degree of similarity we have used a microscopic overlap function~\cite{flenner:2011}
\begin{equation}
w_{mn}(0,t)\equiv \Theta(a-|\mathbf{r}_n(t)-\mathbf{r}_m(0)|),
\end{equation}
where $a=0.3\sigma$ is chosen sufficiently small to enforce single occupancy for hard spheres. Note that this overlap form is different and better suited for the system's geometry than that used in Ref.~\cite{charbonneau:2012c}. For a concentration $c$ of 
pinned particles one therefore has
\begin{equation}
\label{eq_def_overlap}
Q_c(t)\equiv\frac{1}{(1-c)^2N}\left\langle \overline{\sum_{m,n \notin \mathcal B } w_{mn}(0,t)}\right\rangle,
\end{equation}
where the brackets denote an average over equilibrium configurations, the overline represents an average over the different ways to 
pin a fraction $c$ of the particles of a given equilibrium configuration, and the sum is over all unpinned particles, with $\mathcal B$ 
denoting the set of pinned particles.

\begin{figure}
\includegraphics[width=\columnwidth]{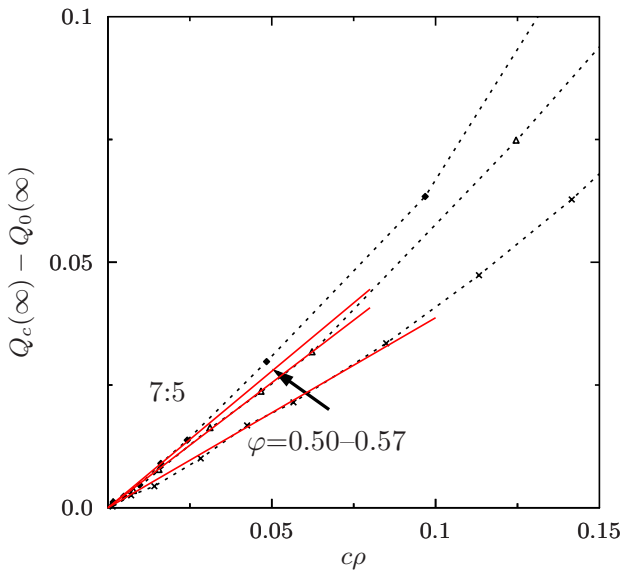}
\caption{(Color online) $Q_c(\infty)-Q_0(\infty)$ versus concentration of pinned particles for the 7:5 binary hard-sphere mixture at $\varphi$=0.50, 0.55, and 0.57. As the packing 
fraction $\varphi$ increases, the growth of the overlap departs more strongly and more rapidly from the linear-response  regime (solid lines).}
\label{fig:pinning_lin}
\end{figure}

The quantity of interest is the long-time limit of $Q_c(t)-Q_0(t)$, ``long-time'' meaning here that the system has relaxed to equilibrium. 
To check that the latter has indeed happened in the observation time, we have monitored the self component of the overlap. For all 
studied densities, the self overlap has indeed completed decayed by the time we measure the asymptotic value of $Q_c(t)-Q_0(t)$. An illustration is provided in Figure~\ref{fig:overlap}. The crossover length between small and large overlap, which provides a proxy for the point-to-set 
correlation length, can then be extracted by locating the rapid decrease of $Q_c(\infty)-Q_0(\infty)$ with $(c\rho)^{-1/3}$. More 
specifically, the length $\xi_\mathrm{p}$ has been defined as the value of the average pinning distance for which the above 
overlap difference falls below $0.4$. Note that the extracted length is not very sensitive to this choice, provided it is intermediate 
between low and high overlap. More methodological details can be found in Ref.~[\onlinecite{charbonneau:2012c}]. It is worth recalling that 
the crossover takes place away from the linear regime in concentration of pinned particles. This regime indeed only contains information on the usual  
static pair correlation function (Fig.~\ref{fig:pinning_lin})~\cite{charbonneau:2012c}.

In addition, we have used Eq.~\eqref{eq_bound_tau} to estimate a lower bound on the growth of static point-to-set correlation lengths. 
From an Arrhenius-like argument for activation volumes~\cite{berthier:2009}, one expects 
$B \propto \beta P$ for hard-sphere fluids~\cite{charbonneau:2012,charbonneau:2012c}, indicating that  the upper 
bound of $\tau_\alpha$ diverges with pressure even in the absence of 
any growing $\xi_{\rm PS}$, as when approaching $T$=0 for an Arrhenius 
temperature dependence. In the low and moderate density fluids, the relaxation time indeed follows 
$\tau_\alpha(P)\simeq \tau_{\alpha,\rm low}(P)=\tau_0\exp(K \beta P)$ with $K$ being a density-independent constant~\cite{footnote:3}.
One then finds that
\begin{equation}
\label{eq_bound-length}
\frac{\xi_{\rm PS}(P)}{\xi_{\rm PS,0}} \gtrsim \left(\frac{\log [\tau_\alpha(P)/\tau_{0}]}{\log[\tau_{\alpha,\rm low}(P)/\tau_0]}\right)^{1/3}\,  ,
\end{equation}
where $\xi_{\rm PS,0}$ is the low-density limit of $\xi_{\rm PS}$ and is related to $K$. The right-hand side of Eq.~\eqref{eq_bound-length} 
thus provides a lower bound for the increase of any static length imposed by the dynamical slowdown.

Finally, note that various measures of the spatial extent of the 
frustrated local tetrahedral order are reported and discussed in detail in Refs.~[\onlinecite{charbonneau:2012,charbonneau:2012c}]. 

\subsection{Dynamical length}
\label{sect:dynamical}
The dynamical relaxation of the fluid structure in the absence of pinning ($c=0$) can be obtained from the microscopic overlap function
\begin{equation}
F_0(t)=\frac{1}{N}\left\langle\sum_{n=1}^{N} w_{nn}(0,t)\right\rangle,
\end{equation}
which is similar to the self-intermediate scattering function $F_s(q,t)$ for a wavevector $q$ near the first peak of the structure factor $S(q)$. 
The structural relaxation time $\tau_\alpha$ can thus be approximated from the $1/e$ decay of $F_0(t)$~\cite{flenner:2011}. The results 
for the two hard-sphere mixtures are shown in Fig.~\ref{fig:relaxtime}.

\begin{figure}
\includegraphics[width=\columnwidth]{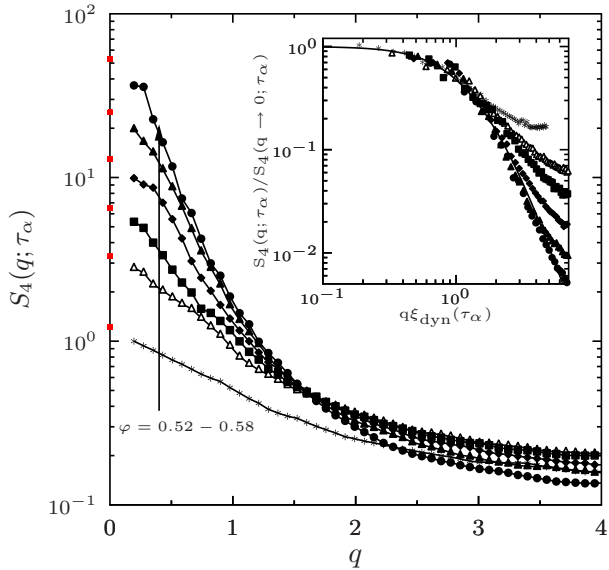}
\caption{(Color online) $S_4(q,\tau_\alpha)$ for the 6:5 mixture at $\varphi$=0.52, 0.55, 0.56, 0.57, 0.575, 0.58. 
(Inset) Collapse of the low-$q$ behavior of $S_4(q,\tau_\alpha)$ using the Ornstein-Zernike form from 
Eq.~\eqref{eq:OZ} (solid line) using the extrapolated intercept values indicated on the $y$ axis in the main panel (empty squares). Note that higher packing fractions follow the form down to lower relative amplitudes.}
\label{fig:S4k}
\end{figure}

Upon slowing down the fluid is known to exhibit fluctuations in particle mobility on a growing spatial range on the timescale 
$\tau_\alpha$~\cite{berthier:2011b}. The size of these regions defines a dynamical length $\xi_{\rm{dyn}}$, which can be 
extracted from the computation of a four-point space-time correlation function~\cite{flenner:2011},
\begin{align}
G_4(r;&\tau_\alpha)=\frac{V}{\langle N_s(\tau_\alpha)\rangle(\langle N_s(\tau_\alpha)\rangle-1)}\nonumber\times\\
&\left\langle\sum_{n\neq m}w_{nn}(0,\tau_\alpha)w_{mm}(0,\tau_\alpha)\delta[\mathbf{r}-\mathbf{r}_{nm}(0)]\right\rangle,
\end{align}
where $N_s(\tau_\alpha)=\sum_n w_{nn}(0,\tau_\alpha)$ is the number of slow particles on the structural relaxation timescale. The 
correlation length $\xi_{\rm{dyn}}=\xi_4(\tau_{\alpha})$ could be obtained from fitting $G_4(r;\tau_\alpha)-G_4(r\rightarrow\infty;\tau_\alpha)$ 
to $\exp(-r/ \xi_{\rm{dyn}})/r$. This direct procedure is, however, numerically difficult because modulations arising from the fluid 
structure are superimposed on the spatial decay of $G_4(r,\tau_\alpha)$. More robustly, we have used the Fourier space version of the function
\begin{align}
S_4(q;\tau_\alpha)=\frac{1}{N}\big[\langle W(\mathbf{q};0,\tau_\alpha)&W(-\mathbf{q};0,\tau_\alpha)\rangle\nonumber\\ 
&-|\langle W(\mathbf{q};0,\tau_\alpha)\rangle|^2\big],
\end{align}
where
\begin{equation}
W(\mathbf{q};0,\tau_\alpha)=\sum_n w_{nn}(0,\tau_\alpha)e^{-i\mathbf{q}\cdot \mathbf{r}_n(0)},
\end{equation}
which measures the structure factor of the regions in the system that remain immobile between times $0$ and $\tau_\alpha$. Fitting the low-$q$ 
result to an Ornstein--Zernike form 
\begin{equation}
\label{eq:OZ}
S_4(q;\tau_\alpha)=\frac{S_4(0;\tau_\alpha)}{1+ (q\, \xi_{\rm{dyn}})^2}
\end{equation}
then provides the dynamical length $\xi_{\rm{dyn}}$.  This 
analysis follows closely that of Ref.~[\onlinecite{flenner:2011}] for the 7:5 binary mixture, 
so we use these published values for this system, 
and only calculate the results for the 6:5 mixture. The raw data for $S_4(q;\tau_\alpha)$ are displayed in Fig.~\ref{fig:S4k}, and 
the inset shows that a reasonable collapse of the low-$q$ regime of $S_4(q,\tau_\alpha)$ is 
obtained by fitting Eq.~\eqref{eq:OZ} to these data.

\section{Results}
\begin{figure*}
\includegraphics[width=1.9\columnwidth]{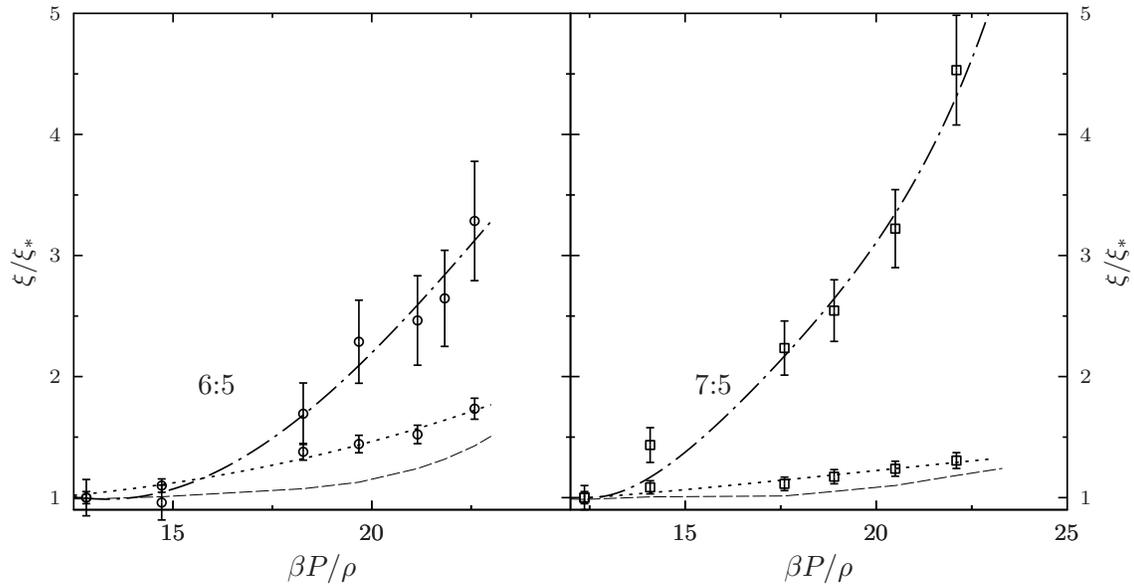}
\caption{For both the 6:5 (left) and the 7:5 (right) binary mixtures, the static (point-to-set) length 
$\xi_{\mathrm{p}}$ (dotted line) grows much slower than the 
dynamical length $\xi_{\rm{dyn}}$ (dash-dotted line) relative to their value at $\varphi_*=0.52$, the onset of nontrivial 
dynamics~\cite{ozawa:2012}. The static length saturates and follows the bound given by Eq.~\eqref{eq_bound-length} (dashed-line). The dynamical 
length $\xi_{\rm{dyn}}$ is taken from Ref.~[\onlinecite{flenner:2011}] for 
the 7:5 mixture and is extracted from fitting the low-wavevector behavior of $S_4(q,\tau_\alpha)$ to Eq.~\eqref{eq:OZ} for the 6:5 mixture. 
Lines are guides for the eye.}
\label{fig:dynlength}
\end{figure*}

For the two glass-forming binary hard-sphere mixtures we find that the point-to-set correlation length $\xi_\mathrm{p}$  
increases but very modestly (by 80\% for the 6:5 mixture and less than 50\% for the 7:5 mixture, see Fig.~\ref{fig:dynlength}) 
for a density range over which the relaxation 
time $\tau_\alpha$ and the diffusivity $D$ change by about 4 orders of magnitude 
(Fig.~\ref{fig:relaxtime}). As already shown in Ref.~[\onlinecite{charbonneau:2012}], 
the structural lengths associated with local order vary 
even less than $\xi_\mathrm{p}$. Note that the bound $\xi_{\rm{PS}}$ given by the right-hand side of Eq.~(\ref{eq_bound-length}) also 
increases quite slowly and is comparable or less than $\xi_{\rm p}$.  (In the presently accessible domain of density, one 
may indeed expect that point-to-set correlation lengths defined either by random 
pinning or by a cavity procedure are of the same order, without presupposing what could happen at yet higher densities.) 
Meanwhile, the dynamical length $\xi_\mathrm{dyn}$ characterizing the spatial extent of heterogeneities in the dynamics 
markedly increases over the density range under study: $\xi_\mathrm{dyn}$ grows by a factor of almost $4$ 
for the 6:5 mixture and $4.5$ for the 7:5 mixture, with no sign of saturation (see also Ref.~[\onlinecite{flenner:2011}]).

These results help answer the first question raised in the introduction. 
The growths of the dynamical and of the static lengths are not systematically correlated as the relaxation 
slows down. They even strongly decorrelate in the dynamical 
regime that is accessible to computer simulations, and to most colloidal experiments, which roughly corresponds to a 
4 order-of-magnitude increase of the relaxation time. The magnitude of this decoupling seems 
to be system dependent~\cite{footnote:1}. 
The divergence of the dynamical and static lengths is more spectacular  in the 7:5 mixture than in the 
more weakly ``frustrated'' 6:5 mixture. For the latter, one could argue that the two 
types of lengths grow at the same pace at low pressure, while the relaxation time increases by, say, 
one order of magnitude. Yet, even in this case, the two quantities eventually unambiguously part ways, in agreement with the results of Refs.~\cite{sausset:2008d,sausset:2010}.

\begin{figure}
\includegraphics[width=\columnwidth]{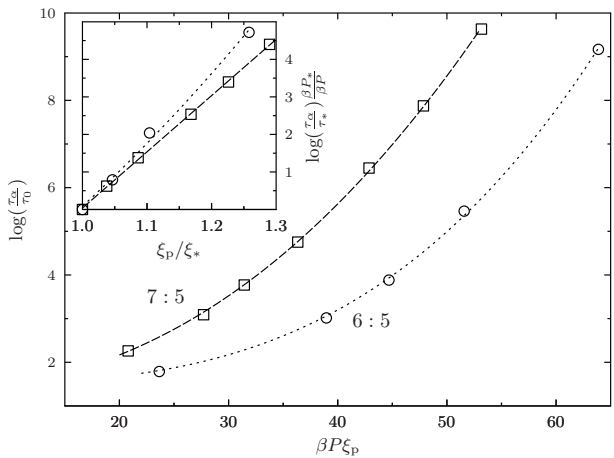}
\caption{Increase of the logarithm of the relaxation time $\log[\tau_{\alpha}(P)/\tau_0]$  
with $\beta P \xi_{\rm p}$ for two binary hard-sphere mixtures (7:5 squares; 6:5 circles). The lines are guides for the eyes that merely 
illustrate the nonlinearity of the dependence. (inset) $\log[\tau_{\alpha}(P)/\tau_*](\beta P/\beta P_*)$ versus $\xi_{\rm p}/\xi_{*}$, where 
time, pressure, and length are normalized by their value at $\varphi_*=0.55$. The lowest density point of the main figure 
is then dropped. Note that because of the implicit offset in the intercept, the roughly linear appearance of 
the plots does not have a simple interpretation in terms of an activated scaling expression with $\psi$.}
\label{fig:fragility}
\end{figure}

We have also investigated a possible correlation between relaxation time and static length for the two hard-sphere binary mixtures. Figuse~\ref{fig:fragility} shows that the observed behavior is quite different from that reported in Ref.~\cite{hocky:2012}, where a 
data collapse for all three studied three-dimensional glass-forming liquids was obtained when using a simple linear 
dependence corresponding to an activated-like scaling expression with $\psi=1$. Here, we find that a linear 
fit clearly does not describe the data and that the two hard-sphere mixtures 
cannot be collapsed onto a unique master curve. As seen in the inset of Fig.~\ref{fig:fragility}, this is true even when 
restricting the analysis to densities above $\varphi=0.55$. (Note that the ``onset'' value above which 
nontrivial glassy dynamics is reported to be around $0.52$~\cite{ozawa:2012}.) It should be stressed that the span 
of relaxation times described in Ref.~\cite{hocky:2012} is rather limited, covering only $1.5$ orders of magnitude.

This second (empirical) finding of a nonuniversal relation between time and length, together with the 
very modest increase of all static lengths potentially associated with the 
slowdown of relaxation, casts doubts on the existence of a general, one-to-one, causal relation between the two quantities. 
In the simulation accessible regime of three-dimensional hard-sphere glass formers, at least, such a relation is not observed.

\section{Conclusion}

By studying two different three-dimensional binary hard-sphere glass formers, we have shown 
that the dynamical length associated with the increasing heterogeneous character of the 
dynamics and the various static lengths that have been put forward to 
explain the collective nature of the dynamical slowdown, be they structural lengths 
associated with local order or point-to-set correlation ones, unambiguously decorrelate 
as density increases. This result is obtained in the dynamical regime that is accessible to 
computer simulations, which covers 4 orders of magnitude in relaxation time and diffusivity. 
All considered lengths increase with density and relaxation time, 
but the dynamical length grows much more rapidly than the static ones.

This finding is sufficient to rule out a general principle tying together the evolutions of 
dynamical and static lengths in glass-forming systems. It is possible however that the 
absence or presence of correlation between the quantities depends on the dynamical 
regime under consideration as well as on the type of material. A strong correlation 
among length scales is expected if glass formers are close enough to a putative 
thermodynamic critical point, whether avoided or unreachable~\cite{tarjus:2011}. 
This phenomenon is what is predicted for instance in weakly frustrated systems, 
as is possibly observed in some two-dimensional systems~\cite{kawasaki:2007,shintani:2008,
watanabe:2008,sausset:2008d,sausset:2010,malins:2012,xu:2012b}, and near a random first-order 
transition to an ideal glass~\cite{lubchenko:2007}.

The link between relaxation time and static correlation length derived  by 
Montanari and Semerdjian~\cite{montanari:2006}, which we have somewhat 
heuristically extended and used in this paper, puts a bound on the contribution that can be 
attributed to a collective or ``cooperative'' activated mechanism driven by the growth 
of a static length scale. We have seen that this contribution stays rather modest in the dynamical range studied. 
In conjunction with the fact that no master curve is
found to collapse the dependence of the relaxation time on the static length for the two 
different hard-sphere mixtures, this result points to the absence of a one-to-one correspondence between 
time and static length in the range under study, and therefore to the absence of a direct 
causal link. Lengths and time all grow in concert but it is impossible to assign them a unique origin 
on the basis of simulation data alone. A couple of factors indeed blur this issue:

\begin{itemize}
\item Several relaxation mechanisms are likely to coexist and entangle in the accessible 
regime, with contributions coming from both cooperativity-driven effects characterized by a 
static length and facilitation or flow/mode-coupling effects characterized by a dynamical length. The signature of a competition 
between different relaxation mechanisms in the same regime has been recently observed in 
related models~\cite{kob:2012,berthier:2012b} (see also Ref.~\cite{flenner:2013}).

\item In a regime where length scales are modest (especially the static ones), various determination methods
may lead to different results, which further obscures the search for a 
causal relation with the structural relaxation slowdown.
\end{itemize}

It is possible that there exists an asymptotic regime in which one mechanism dominates and characteristic 
lengths become very large so that one can causally attribute the dynamical slowdown to the growth of 
a unique typical length associated with  the extent of collective/cooperative behavior driving the dynamics. 
This phenomenon is what several theories that involve a singularity of one form or another predict. Yet, such 
a regime, because of timescale limitation or irreducible frustration or obstacles, appears to be out of reach of (present day) 
simulation studies. Changing the curvature of space to increase the static 
length~\cite{sausset:2008d,sausset:2010} or increasing the dimensionality of space 
to decrease it in the dynamically accessible regime~\cite{eaves:2009,charbonneau:2011,charbonneau:2012b,
charbonneau:2012c,sengupta:2013}, may therefore be 
more productive numerical approaches to understanding the glass problem.

\begin{acknowledgments}
We acknowledge stimulating interactions with L.~Berthier, R.~L.~Jack and D.~R.~Reichman. PC acknowledges NSF support No.~NSF DMR-1055586. 
\end{acknowledgments}


\begin{thebibliography}{75}
\bibitem [{foo({\natexlab{b}})}]{footnote:3}%
  \BibitemOpen
  \href@noop {} {} \bibinfo {note} {To account for a simple
  dependence of the activation volume on the density, one could also consider
  that $B \propto \beta P/\rho$, \textit{i.e.}, that $K\propto1/\rho$ (see
  \textit{e.g.}, Ref.~\cite{berthier:2009}). The modification, however, is not
  significant.}\BibitemShut {Stop}%
\bibitem [{foo({\natexlab{c}})}]{footnote:1}%
  \BibitemOpen
  \href@noop {} {} \bibinfo {note} {We use the terms
  ``decoupling'' or ``decorrelation'' to describe diverging evolutions with
  increasing density and relaxation time because in the fluid regime, at
  sufficiently low density, all length scales are in a sense correlated and
  comparable to the particle size.}\BibitemShut {Stop}%
\end{thebibliography}
%

\end{document}